\UseRawInputEncoding
\documentclass[aps,superscriptaddress,showpacs,nofootinbib,twocolumn]{revtex4}
\usepackage{amsmath}
\usepackage{slashed}
\usepackage[dvips]{color}                 %%% for colours in the graphs
\usepackage{graphicx}
\usepackage{eucal}

\usepackage{mathbbol}

\newcommand{\beq}{\begin{equation}}
\newcommand{\eeq}{\end{equation}}
\newcommand{\bea}{\begin{eqnarray}}
\newcommand{\eea}{\end{eqnarray}}

\usepackage{epsfig}

\begin{document}

\title{Effective Fermion Mass in Relativistic and Non-Relativistic Systems}

\author{Heron Caldas} \email{hcaldas@ufsj.edu.br} \affiliation{Departamento de
  Ci\^{e}ncias Naturais, Universidade Federal de S\~{a}o Jo\~{a}o del Rei,\\
  36301-160, S\~{a}o Jo\~{a}o del Rei, MG, Brazil}

\begin{abstract}
Electrons interact strongly with their environment. The result of these interactions is, most of the time, encoded in an effective mass. In non-relativistic systems, as in condensed matter, the electrons plus interactions form a quasiparticle with an effective mass. From the side of relativistic systems, the fermions also acquire an effective mass due to the interactions with the surrounding medium. We employ a non-perturbative method to calculate the effective mass of relativistic and non-relativistic fermions, in various situations. We find the effective masses up to second order of the iteration method. The results can be of interest in current studies on fermion systems.

\end{abstract}


\maketitle

\section{Introduction}

According to quantum electrodynamics (QED), the bare electron, with charge $e_0$ and mass $m_0$, is surrounded by a virtual cloud of photons. Corrections to these bare quantities have infinite results. However, the observed electron {\it physical} charge $e$ and mass $m$ have finite values. In order to avoid these infinities, one often uses regularization, e.g., cutting off all integrals at large integration momenta.  As a result, the dressed, effective electron charge and mass become finite in the renormalized theory~\cite{Peskin}.

In analogous way, in condensed matter (non-relativistic) bare electrons are dressed with interactions with the many-body system, forming a ``quasiparticle''~\cite{Mattuck}. The non-relativistic quasiparticles we will study here results from the interaction of the electrons with phonons, and consistis of an electron plus a phonon cloud. The basic effect of electron-phonon interaction in metals is the scattering of the electrons by phonons which is an important effect on resistance, and consequently, on transport in metals. One of most familiar consequences of the existence of the electron-phonon interaction is a change in the electron energy and effective mass~\cite{Grimvall}. In several situations the effective mass increases~\cite{Grimvall,Bohm}, decreases~\cite{Krakovsky,Vignale}, and it may even be negative~\cite{Scazza,Cui}.

As is well known, the naive perturbative expansion in powers of the coupling constant breaks down in some situations as, for instance, in theories with spontaneously broken symmetries requiring {\it non-perturbative} resummation schemes to obtain reliable results~\cite{T1,T2,T3}. The second typical case happens in asymptotically free theories where one is not always able to do perturbation theory, since the coupling depends on the energy scale, and at low energies the interaction becomes strong~\cite{MMarino}.

Motivated by these examples, in this work we investigate the effective mass in relativistic and non-relativistic two- and three-dimensional systems, with the help of a non-perturbative resummation approach.

Our main goals in the relativistic framework is to calculate the zero temperature electron self-energy correction to the bare mass $m_0$ in 3D (three space dimensions), and to obtain an effective mass to massless Dirac fermions in graphene, in 2D. In the non-relativistic scenario we aim to calculated the quasiparticle effective mass $m^*$ as a function of the dimensionless coupling constant in the 3D electron-phonon system, both by means of a non-perturbative method.

This paper is organized as follows. In section \ref{section-relat} we present the formalism of our theoretical approach. In this section we also calculate the corrections to the bare mass of relativistic electrons in 3D i.e., in the framework of QED, and to massless Dirac fermions in graphene in 2D. In section \ref{section-nonrelat}, in order to obtain the quasiparticle effective mass we calculate the corrections to non-relativistic electron mass, as well as the polaron ground-state energy and mass, due to electron-phonon interaction. In section \ref{section-DuBois} we comment on the similarities to related non-perturbative approaches. We conclude in section \ref{conclusions} with a discussion.

\section{Relativistic electrons}
\label{section-relat}

\subsection{QED in 3D}

Our starting relativistic field theory to present the formalism of our theoretical approach will be QED in 3D. Quantum electrodynamics, or the relativistic quantum field theory of electrodynamics, can be characterized as a perturbation theory of the electromagnetic quantum vacuum. The ``beauty'' of QED is mainly due to its accurate predictions of physical quantities, such as the Lamb shift of the energy levels of hydrogen. The Lamb shift is a small energy difference between the two energy levels $2S_{1/2}$ and $2P_{1/2}$ of the hydrogen atom, and was of uttermost importance for the development of quantum electrodynamics. Interaction between vacuum energy fluctuations and the hydrogen electron in these different orbitals causes the Lamb shift~\cite{Haken}.

The accurate measurement of the Lamb shift for tightly bound electrons is one of the main subjects of research for highly charged ions. To reach the corresponding accuracy in theoretical calculations one has to take into account all QED corrections of order $\alpha$ and $\alpha^2$ and all orders in $\alpha Z$, where $\alpha$ is the fine structure constant and $Z$ is the charge number of the nucleus~\cite{Lamb1,Lamb2}. 

\subsubsection{QED Model Hamiltonian}

Mathematically, QED is the quantum field theory that describes the interaction between electrons and photons, and is described by the Lagrangian

\begin{eqnarray}
\label{Lag-QED}
{\cal L}_{QED} &=& {\cal L}_{Dirac} + {\cal L}_{Maxwell}  + {\cal L}_{int}\\
\nonumber
&=&\bar \psi ( i  \slashed{\partial} - m_0) \psi - \frac{1}{4} (F_{\mu \nu})^2  - e \bar \psi \gamma^{\mu} \psi A_{\mu},
\end{eqnarray}
where $\psi(x)$ denotes the Dirac spinor field operator, $\bar \psi(x) \equiv \psi(x)^\dagger \gamma_0$, $m_0$ is the electron bare mass, $F_{\mu \nu} = \partial_{\mu} A_{\nu} - \partial_{\nu} A_{\mu} $ is the electromagnetic field tensor, $A_{\mu}$ is the electromagnetic vector potential, and $e = |e|$ is the electron charge.

The interaction term in the QED interaction Hamiltonian is considered small,

\begin{eqnarray}
\label{Ham-QED}
H_{int} &=& \int d^3x~ {\cal H}_{int} =  - \int d^3x~ {\cal L}_{int}\\
\nonumber
&=&  \int d^3x~ [ e \bar \psi \gamma^{\mu} \psi A_{\mu}].
\end{eqnarray}
The smallness of the interaction is due to the fact that the parameter which is relevant for the perturbative expansion is the fine structure constant $\alpha = e^2/4\pi \hbar c \simeq 1/137$~\cite{Michele}.

\subsubsection{Relativistic Fermion Self-Energy in 3D}

\begin{figure}[htb]
  \vspace{0.1cm}
  \epsfig{figure=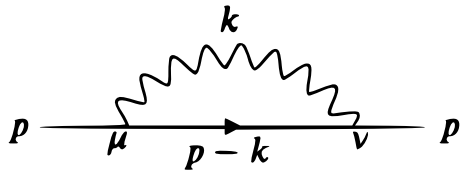,angle=0,width=6cm}
\caption[]{\label{FD1} Lowest order Feynman diagram for the electron-photon contribution to the electron self-energy.}
\end{figure}

We consider now the energy associated with the coupling of an electron to the vacuum electromagnetic field, known as the ``self-energy''. The full propagator of a relativistic electron is given by

\begin{eqnarray}
G = \frac{i}{\slashed{p} -m_0 - \Sigma(p)- i \epsilon},
\label{FProp}
\end{eqnarray}
where $\slashed{p}=p^{\mu} \gamma_{\mu}$, and $\Sigma(p)$ is the sum of all one particle irreducible (1PI) diagrams. $\Sigma(p)$ can be expressed by the diagram in Fig.~(\ref{FD1}). From the above equation, it is very easy to see that the physical mass $m$ is obtained from the pole in the propagator, or $p^2=m^2=(m_0 + \Sigma(p^2=m^2))^2$. Besides, equation~(\ref{FProp}) can also be written as $G^{-1}=G_0^{-1}-\frac{1}{i}\Sigma(p)$, where $G_0$ is the ``free" electron propagator.

In order to consider the above mentioned general situation and with the purpose of studying the electron effective mass, let us formally consider the Schwinger-Dyson (SD) equation for the electron self-energy $\Sigma(p^2)$ depicted in Fig.~\ref{FD1}. Generally, the SD equation has the form

\begin{eqnarray}
\Sigma(p^2)= \int \frac{d^4 k}{(2\pi)^4} F(\Sigma(p^2)) D(k),
\label{S-D}
\end{eqnarray}
where the function $F(\Sigma(k^2))$ depends on the full fermion propagator, and $D(k)$ is the photon propagator. Applying the Feynman diagram rules for QED in $4$ dimensions and in the Feynman gauge ($\xi = 1$) for the diagram in Fig.~\ref{FD1}, we find that the analytic expression for $\Sigma(p^2)$ is given by

\begin{widetext}
\begin{eqnarray}
\Sigma(p^2) = i  \int \frac{d^4 k}{(2 \pi)^4} (- i e \gamma^\mu) \frac{i [ \gamma \cdot (p-k) +m ] }{(p-k)^2 - m^2} (- i e \gamma^\nu) \frac{-i g_{\mu \nu}}{k^2}.
\label{Sigma}
\end{eqnarray}
\end{widetext}
Standard field theory calculations shown in Appendix \ref{ApA} give

\begin{eqnarray}
\Sigma = \frac{3 \alpha}{4 \pi} m \left( \ln \left( \frac{\Lambda^2}{m^2} \right) + \frac{1}{2} \right),
\label{Self1}
\end{eqnarray}
where $\Lambda$ is a momentum cutoff. This result was first obtained in QED by V. Weisskopf~\cite{Weisskopf}. 

From the pole of the full propagator in Eq.~(\ref{FProp}) we obtain,

\begin{eqnarray}
m = m_0 \left[1+ \frac{3 \alpha}{4 \pi}  \left( \ln \left( \frac{\Lambda^2}{m_0^2} \right) + \frac{1}{2} \right) \right],
\label{Mass1}
\end{eqnarray}
where the mass $m$ in the self-energy has been replaced by the bare mass $m_0$. This means that the self-energy was calculated to lowest order. As we will see below, this result corresponds to an approximation equivalent to the ``first-order'' iteration in the evaluation of the self-energy, i.e., $m=m_0 + \Sigma^0 =m_0 + \Sigma(m_0)$.

Excepting gravitation, all theories describing fundamental interactions, as is the case of QED, which describes all the phenomena involving electrically charged carriers interacting by means of exchange of photons, are renormalizable theories. Therefore, the issue of renormalization of a theory is of fundamental importance.

In order to proceed with the mass renormalization, we add to the original Hamiltonian the counterterm $-\tilde m \bar \psi  \psi = - \frac{3 \alpha}{4 \pi} m_0  \ln \left( \frac{\Lambda^2}{m_0^2} \right)  \bar \psi  \psi$, such that

\begin{eqnarray}
m = m_0 \left[1+ \frac{3 \alpha}{8 \pi}  \right].
\label{Mass2}
\end{eqnarray}

\subsubsection{The Non-perturbative Fermion Self-Energy: The MSCR}

The Schwinger-Dyson equation in Eq.~(\ref{S-D}) is a transcendental equation. Because the unknown function is under the integral, the integration cannot be performed analytically.

Nevertheless, one can try to solve the SD equation by means of some approximation or iterative method.

Here we apply the Modified Self-Consistent Resummation (MSCR) in order to implement a resummation of infinity sub-set of diagrams~\cite{Caldas1,Caldas2,Caldas3}. The method consists in calculating the mass through iteration, assuring the mass renormalization in each step. 

The divergences from the integrals computed in a present step are exactly canceled by the counter-terms that are necessary to be introduced at the previous order of the recalculation (resummation) of the self-energy to carry the renormalization. This renormalization method has been developed, originally in the context of low energy quantum chromodynamics (QCD), in Refs.~\cite{Caldas1,Caldas2,Caldas3} for the resummation method employed here and has been given the name MSCR. For the purposes of this work, we concentrate on the finite terms and refer the reader to the above cited works for a detailed treatment of the renormalization procedure.

For all $n \geq 1$ we have 

\begin{eqnarray}
m_n=m_0 + \Sigma(m_{n-1}),
\label{Method1}
\end{eqnarray}
with $m_{n=0} \equiv m_0$. 

Strictly speaking, the iteration should continue until one obtains the equivalent of a ``gap equation'' for $m$

\begin{eqnarray}
m =m_0 + \Sigma(m).
\label{Method2}
\end{eqnarray}

Another way to say this is in terms of the order $n$ of the iteration; one should iterate until some $n^*$ such that a stabilization occurs, i.e., $m_{n^*} - m_{n^*-1} \approx 0$, or $\Sigma(m_{n^*}) \approx \Sigma(m_{n^*-1})$.

For the present case we follow the steps:\\

I) Calculate the corrected mass to lowest order

\begin{eqnarray}
m_1 &=& m_0 + \Sigma(m_0)\\
\nonumber
 &=& m_0 \left[1+ \frac{3 \alpha}{4 \pi}  \left( \ln \left( \frac{\Lambda^2}{m_0^2} \right) + \frac{1}{2} \right) \right].
\label{MassM1}
\end{eqnarray}
\\
II) Define the renormalized mass to lowest order (i.e., Eq.~(\ref{Mass2})) as 

\begin{eqnarray}
m_{1r}=m_0 \left[1+ \frac{3 \alpha}{8 \pi}  \right].
\label{MassM1}
\end{eqnarray}
\\
V) Calculate again the self-energy in Eq.~(\ref{Self1})  with $m_{1r}$, which gives

\begin{eqnarray}
m_2 &=& m_0 + \Sigma(m_{1r})\\
\nonumber
&=& m_0 + \frac{3 \alpha}{8 \pi} m_{1r} + \frac{3 \alpha}{8 \pi} m_{1r} \ln \left( \frac{\Lambda^2}{m_{1r}^2} \right).
\end{eqnarray}
\\
IV) Renormalize the mass, absorbing only the divergente piece as before, which yields

\begin{eqnarray}
m_{2r} = m_0\left[1 +  \frac{3 \alpha}{8 \pi} + \left( \frac{3 \alpha}{8 \pi}  \right)^2 \right].
\label{MassM2-ren}
\end{eqnarray}

Now, instead of absorbing the divergent part of $m_1$ before calculating $m_2$, we will keep the full expression of $m_1$ in order to compare with a $\alpha^2$ result from a perturbative calculation.\\
\\
V) Calculate again the self-energy in Eq.~(\ref{Self1})  with $m_{1}$, which gives

\begin{eqnarray}
m_2 &=& m_0 + \Sigma(m_1)\\
\nonumber
&=&m_0 \left[1+ \frac{3 \alpha}{4 \pi}  \left( \ln \left( \frac{\Lambda^2}{m_0^2} \right) + \frac{1}{2} \right) \right]\\
\nonumber
&+&m_0 \left( \frac{ \alpha}{4 \pi} \right)^2 \left[ 9 \ln^2 \left( \frac{\Lambda^2}{m_0^2} \right)+ 9\ln \left( \frac{\Lambda^2}{m_0^2} \right) + \frac{9}{4}  \right]\\
\nonumber
&+& F(\alpha),
\label{Mass3}
\end{eqnarray}
where the function $F(\alpha)$ is of superior order in $\alpha$,

\begin{eqnarray}
\label{F}
F(\alpha)&\equiv& \frac{3 \alpha}{4 \pi} m_1 \ln \left( \frac{m_0^2}{m_1^2} \right)\\
\nonumber
&=& -  \frac{3 \alpha}{4 \pi} m_0 \left[1+ \frac{3 \alpha}{4 \pi}  \left( \ln \left( \frac{\Lambda^2}{m_0^2} \right) + \frac{1}{2} \right) \right]\\
\nonumber
&\times&  \ln \left[1+ \frac{3 \alpha}{4 \pi}  \left( \ln \left( \frac{\Lambda^2}{m_0^2} \right) + \frac{1}{2} \right) \right]^2.
\end{eqnarray}

The result of order $\alpha^2$ of the perturbative expansion, i.e., which comes from the computation of Feynman graphs for the second-order QED correction, is given by~\cite{Leonti,Leonti2}

\begin{eqnarray} 
\delta m &=& m_0 \left( \frac{ \alpha}{4 \pi} \right)^2  \left\{ \frac{15}{2} \ln^2 \left( \frac{\Lambda^2}{m_0^2} \right) + 3\ln \left( \frac{\Lambda^2}{m_0^2} \right) \right. \\ 
\nonumber
&+& \left.  \frac{\pi^2}{2} - \frac{63}{4} \right\}.
\label{Leonti}
\end{eqnarray}
Notice that the terms of order $\alpha^2$ from the non-perturbative calculation does not agree with accuracy to the perturbative ones. As is widely known, perturbative QED has been proved to be a {\it consecrated} theory with predictions that has been tested over the years with very high precision. See, for instance, Refs.~\cite{Hanneke} and \cite{Sturm}. However, it is worth to emphasize that non-perturbative calculations are relevant (or even inevitable~\cite{Bakulev}) in situations where perturbative calculations break-down or are not applicable.

\subsection{QED in 2D}

\subsubsection{Relativistic fermions with zero effective mass}

Graphene is a two-dimensional (2D) allotrope of carbon, made out of carbon atoms arranged on a honeycomb structure. The electrons in graphene are elegantly described by (a formally equivalent to) the Dirac Hamiltonian around the Dirac point \footnote{The Dirac points are the contact points between the valence and conduction bands, which makes undoped graphene a zero-gap material.} $K$ (or $K'$), that in momentum space reads~\cite{Kotov},

\begin{equation}
\label{DiracGraphene}
   H= \hbar v_F
  \left[ {\begin{array}{cc}
   0 & q_x + i q_y \\
    q_x - i q_y & 0 \\
  \end{array} } \right] = \hat{\sigma} \boldsymbol{\cdot} \vec{q},
\end{equation}
where the components of the operator $\hat{\sigma}$ are the usual Pauli matrices. The Hamiltonian in Eq.~(\ref{DiracGraphene}) is exactly that of an ultra-relativistic (or massless) particle of spin $1/2$ (such as the neutrino), with the velocity of light $c$ replaced by the Fermi velocity $v_F=1/300 c$. The eigenvalues of $\rm{H}$ are given by

\begin{equation}
\label{EVDE}
\epsilon(q)= \pm \hbar v_F |q|,
\end{equation}
which means that electrons propagating through graphene's honeycomb lattice lose their mass, behaving as quasi-particles that are described by a 2D analogue of the Dirac equation, rather than the (non-relativistic) Schrodinger equation for spin-$1/2$ particles, as we will see in the next Section.

\subsubsection{Relativistic fermions with a finite effective mass}

One can define an effective mass $m^*$ in the standard way as

\begin{equation}
\label{meffD1}
m^*= \frac{\hbar k_F}{v_F} = \frac{\hbar (\pi n)^{1/2}}{v_F},
\end{equation}
where $ k_F=(\pi n)^{1/2}$ is the 2D Fermi radius, $n=$ number density of (additional) electrons = number of additional electrons/unit area, which means that this refers to extrinsic (or doped) graphene, as in any real system, with a finite chemical potential (and consequently finite carrier density)~\cite{Das}. Indeed, experiments have observed an effective cyclotron mass $m_c$ that depends on the electronic density as $n^{1/2}$~\cite{Novoselov,Zhang}. This provides a direct evidence for the existence of massless Dirac quasiparticles in graphene.

\subsubsection{Relativistic fermions with an effective mass from the MSCR}

It has been obtained recently the renormalized quasiparticle velocity in graphene up to second-order of the non-perturbative iteration within the MSCR, whose expressions reads~\cite{PhysB},

\bea
%v_{F2}(q) &=&v_F+v_{F1}(q)\frac{\alpha}{4}\log(\Lambda/|\vec{q}|) \to \nn\\
\frac{v_{F2}(q)}{v_{F}}&=&1+  \frac{\alpha}{4}\log(\Lambda/|\vec{q}|) + \left( \frac{\alpha}{4} \right)^2\left[  \log(\Lambda/|\vec{q}|)\right]^2,
\label{vel2}
\eea
where $\alpha$ is an effective fine structure constant for graphene. Making the replacement~\cite{Das} $|\vec{q}| \to q_F=(\pi n)^{1/2}$, where $\Lambda=(\pi n_c)^{1/2}$ is a momentum cutoff, with a fixed reference density $n_c=10^{15}cm^{-2}$, and substituting these in the above equation, we find

\bea
\frac{v_{F2}(n)}{v_{F}}&=&1+  \frac{\alpha}{8}\log \left(\frac{n_c}{n} \right) + \frac{1}{4}\left( \frac{\alpha}{4} \right)^2\left[\log \left(\frac{n_c}{n} \right) \right]^2.
\label{vel-n}
\eea

Then we define an effective mass $m_{eff}$ as

\begin{equation}
\label{meffD2}
m_{eff}= \frac{\hbar k_F}{v_F^*},
\end{equation}
where $\hbar k_F=m_0 v_F$, with $m_0=9.11\times10^{-31} \rm{kg}$ being the free electron mass, and $v_F^* \equiv v_{F2}(n)$ is given by Eq.~(\ref{vel-n}), which yields

\begin{equation}
\label{meffD3}
m_{eff}= \frac{m_0}{1+  \frac{\alpha}{8}\log \left(\frac{n_c}{n} \right) + \frac{1}{4}\left( \frac{\alpha}{4} \right)^2\left[\log \left(\frac{n_c}{n} \right) \right]^2}.
\end{equation}

\begin{figure}[htb]
  \vspace{0.1cm}
  \epsfig{figure=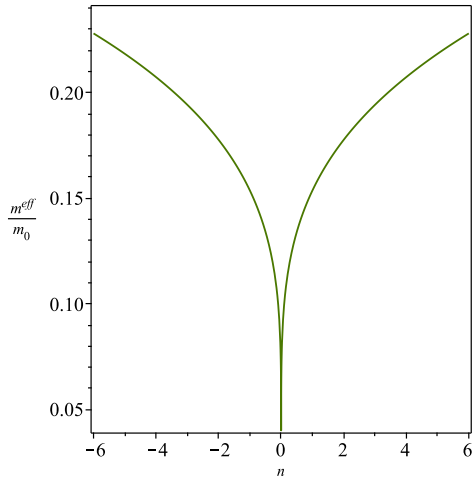,angle=0,width=6.5cm}
\caption[]{\label{EffecMassGraphene} Non-perturbative effective mass of the Dirac fermions in graphene as a function of the carrier density $n$ (in units of $10^{12}cm^{-2}$).}
\end{figure}

In Fig.~\ref{EffecMassGraphene} we show the behavior of the ratio of effective mass to the free electron mass, $m_{eff}/m_0$ from Eq.~(\ref{meffD3}), for $\alpha=2.2$ (for graphene suspended in vacuum~\cite{Elias}), as a function of the density $n$. Notice that $m_{eff}$ presents the same qualitative behavior of $m^*$ from Eq.~(\ref{meffD1}). The two effective masses also agree in the limiting cases (notice that the $v_F$ appearing in Eq.~(\ref{meffD1}) is $v_F= \hbar k_F(n_c)/m_0=\hbar \Lambda/m_0=10^6m/s$): for $n=0$, $m_{eff}=m^*=0$, and for $n=n_c$, $m_{eff}=m^*=m_0$.

\section{Non-relativistic electrons}
\label{section-nonrelat}

Non-relativistic electrons are characterized by their parabolic dispersion 

\begin{eqnarray}
\xi_{k} = \frac{k^2}{2m} - \mu,
\label{Energy1}
\end{eqnarray}
where $\mu$ is the chemical potential. As we will see next, the single particle $m$ will also be modified (as happens with the relativistic electrons) to the effective mass $m^*$, resulting from the electron-phonon interaction.

\subsubsection{Fr\"{o}hlich Polarons}

Fr\"{o}hlich realized that the electron-phonon interaction is closely analogous to the electron-photon interaction of quantum electrodynamics~\cite{Piers}, that we have studied in the previous Section. He found that an electron surrounded by a cloud of (virtual) phonons, i.e., a polaron, has its energy given by

\begin{eqnarray}
E_{k} = -\alpha_{Fep} \hbar \omega_{LO} + \frac{k^2}{2m^*},
\label{Polaron1}
\end{eqnarray}
where 

\begin{eqnarray}
\alpha_{Fep}=\frac{e^2}{\hbar \bar \epsilon}\sqrt{\frac{m_e}{2 \hbar \omega_{LO}}},
\label{e-f}
\end{eqnarray}
is the dimensionless Fr\"{o}hlich electron-phonon coupling constant, $\omega_{LO}$ is the frequency of the longitudinal optical branch of the phonon field, $m_e$ is the free electron mass, and

\begin{eqnarray}
\label{Polaron2}
m^*=m_e \left(1+\frac{\alpha_{Fep}}{6} \right).
\end{eqnarray}
In the definition of $\alpha_{Fep}$,

\begin{eqnarray}
\frac{1}{ \bar \epsilon} =  \frac{1}{ \epsilon_{\infty}} - \frac{1}{ \epsilon_{0}},
\nonumber
\label{Polaron3}
\end{eqnarray}
where $\epsilon_{\infty}$ and $\epsilon_{0}$ are, respectively, the high-frequency and the static dielectric constants of the polar crystal. We remark that the result in Eq.~(\ref{Polaron2}) is expected to be valid only for small $\alpha_{Fep}$~\cite{Frohlich}.

\subsubsection{Electron Effective-Mass from Feynman Diagrams}

The main effect of the electron-phonon interaction on electron propagation, is determined by the electronÐphonon self-energy (SE) $\Sigma(p)$, such that their (free) Green's function

\begin{eqnarray}
\mathcal{G}_{0}(\omega)=\frac{1}{\omega-\epsilon_{\vec{k} }+ i \delta \rm{sign}(\omega)},
\label{G0}
\end{eqnarray}
is modified to

\begin{eqnarray}
\mathcal{G}(\omega)=\frac{1}{\omega-\epsilon_{\vec{k}} - \Sigma(\vec{k}, \omega)},
\label{G0-1}
\end{eqnarray}
where $\Sigma(\vec{k}, \omega)$ is the self-energy, and $\delta$ in Eq.~(\ref{G0}) is to be understood as a positive infinitesimal. In general, the SE can be separated in real and imaginary parts

\begin{eqnarray}
\mathcal{G}(\vec{k},\omega)=\frac{1}{\omega-[\epsilon_{\vec{k} } +\text{Re} \Sigma(\vec{k}, \omega)] - i \text{Im}\Sigma(\vec{k}, \omega)}.
\label{G0-2}
\end{eqnarray}
The renormalized Fermi momentum of the fermions is defined by the condition that the real part of the energy vanishes $\epsilon_{\tilde k_{F }} +\text{Re} \Sigma(\tilde k_{F }, 0)=0$. From Eq.~(\ref{G0-1}) the renormalized spectrum $\tilde\epsilon_{ \vec{k}}$ is a solution of the equation $\omega=\tilde\epsilon_{ \vec{k}}$, or

\begin{eqnarray}
\tilde\epsilon_{ \vec{k}} - \epsilon_{\vec{k} } - \Sigma(\vec{k},\tilde\epsilon_{ \vec{k}})=0.
\label{Expan-2}
\end{eqnarray}
Thus, at small energies and for $k$ close to the pole $\tilde\epsilon_{ \vec{k}}$, we can expand $\mathcal{G}(\vec{k},\omega)$, to obtain~\cite{Daniel}

\begin{eqnarray}
\nonumber
\mathcal{G}(\vec{k},\omega)&=&\frac{1}{\omega-\epsilon_{\vec{k} }- \Sigma(\vec{k},\tilde\epsilon_{ \vec{k}}) -\frac{ \partial\Sigma(\vec{k}, \omega)}{\partial \omega}|_{\omega=\tilde\epsilon_{ \vec{k}}}(\omega-\tilde\epsilon_{ \vec{k}}) }\\
\nonumber
&=& \frac{1}{\omega- \tilde\epsilon_{ \vec{k}} -\frac{ \partial\Sigma(\vec{k}, \omega)}{\partial \omega}|_{\omega=\tilde\epsilon_{ \vec{k}}}(\omega-\tilde\epsilon_{ \vec{k}}) }\\
\nonumber
&=& \frac{1}{(\omega-\tilde\epsilon_{ \vec{k}})} \frac{1}{1-\frac{ \partial\Sigma(\vec{k}, \omega)}{\partial \omega}|_{\omega=\tilde\epsilon_{ \vec{k}}}}\\
&=& \frac{Z_k}{\omega-\tilde\epsilon_{ \vec{k}}},
\label{Expan-1}
\end{eqnarray}
where $Z_k$ is the pole strength, given by

\begin{eqnarray}
Z_k=  \frac{1}{1-\frac{ \partial\Sigma(\vec{k}, \omega)}{\partial \omega}|_{\omega=\tilde\epsilon_{ \vec{k}}}}.
\label{Pole-1}
\end{eqnarray}
From Eq.~(\ref{Expan-2}) it is assumed the following form for the renormalized spectrum

\begin{eqnarray}
\tilde\epsilon_{ \vec{k}} = \frac{k^2}{2m^*}.
\label{Pole-1}
\end{eqnarray}
Then,

\begin{eqnarray}
\frac{1}{2 m^*} &=& \frac{\partial \tilde\epsilon_{ \vec{k}}}{\partial k^2} = \frac{\partial }{\partial k^2} \left(\epsilon_{\vec{k} } + \Sigma(\vec{k},\tilde\epsilon_{ \vec{k}}) \right)\\
\nonumber
&=& \frac{1}{2 m}  + \frac{\partial  \Sigma(\vec{k},\tilde\epsilon_{ \vec{k}})}{\partial k^2} + \frac{\partial  \Sigma(\vec{k},\tilde\epsilon_{ \vec{k}})}{\partial \tilde\epsilon_{ \vec{k}}} \frac{\partial \tilde\epsilon_{ \vec{k}}}{\partial k^2} \\
\nonumber
&=& \frac{1}{2 m}  + \frac{1}{2 m} \frac{\partial  \Sigma(\vec{k},\tilde\epsilon_{ \vec{k}})}{\partial (k^2/2m)} +  \frac{\partial  \Sigma(\vec{k},\tilde\epsilon_{ \vec{k}})}{\partial  \omega}|_{\omega = \tilde\epsilon_{ \vec{k}}} \frac{\partial \tilde\epsilon_{ \vec{k}}}{\partial k^2}.
\label{NewSpec-1}
\end{eqnarray}
Collecting the $m$ and $m^*$ terms we have

\begin{eqnarray}
\frac{1}{ m^*} \left( 1- \frac{\partial  \Sigma(\vec{k},\omega)}{\partial  \omega}|_{\omega = \tilde\epsilon_{ \vec{k}}} \right) = \frac{1}{ m} \left(1 +  \frac{\partial  \Sigma(\vec{k},\omega)}{\partial \epsilon_{ \vec{k}}} \right),
\label{NewSpec-2}
\end{eqnarray}
that finally gives

\begin{eqnarray}
\frac{m^*}{m} = \frac{1}{Z_k} \frac{1}{1+\frac{ \partial\Sigma(\vec{k}, \omega)}{\partial \epsilon_{\vec{k} }}},
\label{EffectiveM-1}
\end{eqnarray}
which defines the effective mass.

In the situations where the interaction leading to mass renormalization is carried out by low-energy excitations, i.e. by electron-phonon interaction with $\omega_{ph} \ll \epsilon_F$, one can neglect the momentum dependence of $\Sigma(p,\omega)$. For this particular case of electron-phonon interaction, one would have

\begin{eqnarray}
\frac{m^*}{m} = \frac{1}{Z_k} \equiv 1 + \lambda,
\label{NewSpec-3}
\end{eqnarray}
where $\lambda = -\frac{ \partial\Sigma(\vec{k}, \omega)}{\partial \omega}|_{\omega=\tilde\epsilon_{ \vec{k}}}$.

\begin{figure}[htb]
  \vspace{0.1cm}
  \epsfig{figure=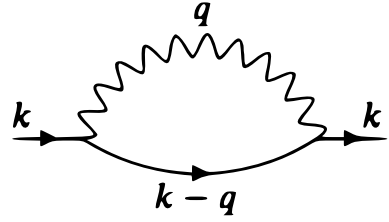,angle=0,width=5cm}
\caption[]{\label{FD2} Lowest order Feynman diagram for the electron-phonon contribution to the non-relativistic electron self-energy. This describes the emission and absorption of a (virtual) phonon.}
\end{figure}

The SE $\Sigma(\vec{k}, \omega)$ can be represented by the diagram in Fig.~(\ref{FD2}). The leading-order Feynman diagram for the self-energy is given by~\cite{Piers}

\begin{eqnarray}
\label{Sigma-NonRelat-1}
\Sigma(k,i\omega_n) &=& \sum_q (i g_q)^2 {\cal G}_0 (k-q) {\cal D}(q)\\
\nonumber
 &=& - T \sum_q  g_q^2 \frac{1}{i \omega_{n} - i \nu_{n} - \epsilon_{ \bf{k} - \bf{q} } } \frac{ 2 \omega_{\bf{q}} }{(i \omega_n)^2 - \omega_{\bf{q}}^2},
\end{eqnarray}
where $g_q$ is the electron-phonon coupling, $\omega_{\bf{q}}$ is the phonon frequency, and ${\cal G}_0 (k-q)$ and ${\cal D}(q)$ are the fermion and boson propagators, respectively. Summation of the frequency $\nu_n$ in Eq.~(\ref{Sigma-NonRelat-1}) gives~\cite{Fetter,Mahan,Piers}

\begin{eqnarray}
\label{Sigma-NonRelat-2}
\Sigma({\bf k},z) =  \sum_q  g_q^2 \left[ \frac{1+ n_{{\bf q}} - f_{{\bf k}-{\bf q}}}{z - ( \epsilon_{ {\bf k} - {\bf q} } + \omega_{{\bf q}}) } + \frac{n_{{\bf q}} + f_{{\bf k}-{\bf q}}}{z - ( \epsilon_{ {\bf k} - {\bf q} } - \omega_{{\bf q}}) } \right],
\end{eqnarray}
where $n_{{\bf q}}$ and $f_{{\bf k}-{\bf q}}$ are the Bose and Fermi functions, respectively. 

Below we will follow from Eq.~(\ref{Sigma-NonRelat-2}) to investigate the ground-state energy and effective mass of the Fr\"{o}hlich polaron.

\subsubsection{The $3D$ Polaron Ground-State Energy and Effective Mass}

The Fr\"{o}hlich polaron ground-state energy and effective mass in Eqs.~(\ref{Polaron1}) and (\ref{Polaron2}), respectively, can be derived from the Fr\"{o}hlich Hamiltonian, which describes the interaction between a single electron and {\it longitudinal optical} (LO) phonon modes~\cite{Frohlich},

\begin{eqnarray}
\label{Ham-Frolich}
H_{F}= H_{el} + H_{ph} + H_{el-ph},
\end{eqnarray}
where the band electrons (noninteracting), phonons and electron-phonon interaction Hamiltonians are given, respectively, by

\begin{eqnarray}
\label{Ham-El}
H_{el} = \sum_{{\bf k},\sigma} \epsilon(k) c_{{\bf k},\sigma}^{\dagger} c_{{\bf k},\sigma},
\end{eqnarray}

\begin{eqnarray}
\label{Ham-Ph}
H_{ph} = \sum_{{\bf q}} \hbar \omega_{{\bf q}} b_{{\bf q}}^{\dagger} b_{{\bf q}},
\end{eqnarray}
and

\begin{eqnarray}
\label{Ham-El-Ph}
H_{el-ph} = \sum_{{\bf k, q}} g_q ( b_{{\bf -q}}^{\dagger} + b_{{\bf q}})c_{{\bf k+q},\sigma}^{\dagger} c_{{\bf k},\sigma},
\end{eqnarray}
where $\epsilon(k)=k^2/2m$, $c_{{\bf k},\sigma}^{\dagger}$ and $c_{{\bf k},\sigma}$ are the electron creation and annihilation operators, $\hbar \omega_{{\bf q}}$ is the dispersion for acoustic phonons, $b_{{\bf q}}^{\dagger}$, and $b_{{\bf q}}$ are the creation operators of the phonon modes, and the vertex $g_q \propto q V(q)$, with $V(q)\propto1/q^2$ the Coulomb potential. The $q^{-2}$ dependence in $V(q)$ comes from the 3D Fourier transform of the (bare) Coulomb potential $\propto 1/r$ of the ionic crystal. For small momentum $q$, i.e., for $q \to 0$, $\hbar \omega_{{\bf q}} \propto q$.

The electron self-energy expression is the same as the one in Eq.~(\ref{Sigma-NonRelat-2}), where $g_q^2 = \frac{M_0^2}{v q^2}$, with $M_0^2 = \frac{4 \pi \alpha \hbar (\hbar \omega_0)^{\frac{3}{2}}}{\sqrt{2m}}$, and $v$ is the volume~\cite{Mahan}. Then we get

\begin{eqnarray}
\Sigma({\bf k}, i k_n) &=&  \frac{M_0^2}{v} \sum_q  \frac{1}{q^2} \\
\nonumber
&\times& \left[ \frac{1+ n_{{\bf q}} - f_{{\bf k} + {\bf q}}}{i k_n - ( \epsilon_{ {\bf k} + {\bf q} } + \omega_{{\bf q}}) } + \frac{n_{{\bf q}} + f_{{\bf k} + {\bf q}}}{i k_n - ( \epsilon_{ {\bf k} + {\bf q} } - \omega_{{\bf q}}) } \right].
\label{Frohlich-1}
\end{eqnarray}
At zero temperature $n_{{\bf q}}=0$ and considering only one electron, $f_{{\bf k} + {\bf q}}=0$. For completeness, we show in Appendix \ref{ApD} the usual evaluation of $\Sigma({\bf k})$ for $f_{{\bf k} + {\bf q}}$ number of fermions.  Setting $i k_n = E - \mu + i \delta$ in the infinity volume ($v \to \infty$), the real part of the retarded self-energy in 3D is

\begin{eqnarray}
{\rm Re}\Sigma({\bf k}, E) &=&  \frac{M_0^2}{(2 \pi)^3} \int  \frac{d^3 q}{q^2}  \frac{1}{ E - \omega_0 - \frac{( {\bf k} + {\bf q} )^2}{2m}} .
\label{Frohlich-2}
\end{eqnarray}
Evaluating the above SE by making the {\it on-shell approximation}~\cite{Giuliani}, which consists of setting $E = \epsilon_k$ in the integral, one finds (see Appendix \ref{ApC})

\begin{eqnarray}
\Sigma({\bf k}) &=&  - \alpha \omega_0  \left[ 1 + \frac{1}{6}\frac{\epsilon_k}{\omega_0} + {\cal O}\left( \frac{\epsilon_k}{\omega_0}\right)^2 \right].
\label{Frohlich-3}
\end{eqnarray}
The ground-state energy is found through the (pole) equation

\begin{eqnarray}
\label{Frohlich-4}
E_k &=& \epsilon_k + \Sigma({\bf k})\\
\nonumber
&=& \epsilon_k  - \alpha \omega_0  \left[ 1 + \frac{1}{6}\frac{\epsilon_k}{\omega_0}  \right]\\
\nonumber
&=& - \alpha \omega_0 + \epsilon_k \left[ 1- \frac{\alpha}{6} \right]\\
\nonumber
&=& - \alpha \omega_0 + \frac{k^2}{2m^*},
\end{eqnarray}
with $m^*$ the effective mass, 

\begin{eqnarray}
m^* = m \frac{1}{1-\frac{\alpha}{6}} \sim m\left(1+\frac{\alpha}{6} \right),
\label{Frohlich-5}
\end{eqnarray}
where $\alpha = \alpha(m)$ is given by Eq.~(\ref{e-f}).

Now, if an attentive reader wants to find the effective mass through Eqs.~(\ref{EffectiveM-1}) and (\ref{Frohlich-3}), it will be found

\begin{eqnarray}
m^* &=& m\left[ \frac{1}{1-\frac{\alpha}{6}} + \frac{\alpha}{1-\frac{\alpha}{6}} \right]\\
\nonumber
&=& m \frac{1}{1-\frac{\alpha}{6}} + {\cal O}\left(\frac{\alpha}{1-\frac{\alpha}{6}} \right),
\label{Frohlich-5-2}
\end{eqnarray}
since $\alpha \ll 1$. Thus, the above result agrees with the one in Eq.~(\ref{Frohlich-5}).

\subsubsection{The Non-perturbative Polaron Ground-State Energy and Effective Mass}

We now apply the MSCR to the polaron energy and effective mass. For simplicity, as we will go only to the second term of the series of iterations, we will retain terms up to $\alpha^2$. To do this, we define the $E_k$ and $m^*$ found above as $E_{k1} = \epsilon_{k1}$, and $m^*=m_1$, and reevaluate equation~(\ref{Frohlich-4}),

\begin{eqnarray}
E_k &=& \epsilon_{k} + \Sigma(\epsilon_{k1})\\
\nonumber
&=& \epsilon_{k}  - \alpha_1 \omega_0  \left[ 1 + \frac{1}{6}\frac{\epsilon_{k1}}{\omega_0}  \right]\\
\nonumber
&=& - \left(1 -\frac{\alpha}{6}\right)\alpha_1\omega_0 + \epsilon_{k}\left[ 1- \frac{\alpha_1}{6}\left(1 -\frac{\alpha}{6}\right) \right]\\
\nonumber
&=& - \alpha_2 \omega_0 + \frac{k^2}{2m_2} \equiv \epsilon_{k2},
\label{Frohlich-6}
\end{eqnarray}
where $\alpha_2 = \left(1 -\frac{\alpha}{6}\right)\alpha_1$, and $\alpha_1 = \alpha(m_1)$, which is given by

\begin{eqnarray}
\alpha_1=\frac{e^2}{\hbar \bar \epsilon}\sqrt{\frac{m_1}{2 \hbar \omega_{0}}},
\label{Frohlich-7}
\end{eqnarray}
such that the effective mass is now

\begin{eqnarray}
m_2 = m \left[1- \frac{\alpha_1}{6} \left(1- \frac{\alpha}{6} \right) \right]^{-1}.
\label{Frohlich-8}
\end{eqnarray}
where $m$ is the bare mass. The highly nonlinear character of the non-perturbative resummation method is evident. 

Now we can find the polaron ground-state energy $E_{02}$ and effective mass $m_2$ up to order $\alpha^3$. With Eq.~(\ref{Frohlich-7}) we obtain $\alpha_1$, and then $\alpha_2$

\begin{eqnarray}
\alpha_1=\alpha + \frac{\alpha^2}{12}  + {\cal O}(\alpha^3),
\label{Frohlich-9}
\end{eqnarray}
which gives

\begin{eqnarray}
\alpha_2&=&\left(1 -\frac{\alpha}{6}\right)\alpha_1\\
\nonumber
&=&  \alpha - \frac{\alpha^2}{12}   + {\cal O}(\alpha^3).
\label{Frohlich-10}
\end{eqnarray}
Then we find

\begin{eqnarray}
E_{02} &=& -  \omega_0 \left[ \alpha - \frac{\alpha^2}{12}  + {\cal O}(\alpha^3) \right]\\
\nonumber
&=& -  \omega_0 \left[ \alpha - 0.083 \alpha^2  + {\cal O}(\alpha^3) \right]
\label{Frohlich-11}
\end{eqnarray}

\begin{eqnarray}
m_2 &=& m \left[1 - \frac{\alpha}{6} + \frac{\alpha^2}{72}  + {\cal O}(\alpha^3) \right]^{-1}\\
\nonumber
&=& m \left[1 - \frac{\alpha}{6} + 0.01389 \alpha^2  + {\cal O}(\alpha^3) \right]^{-1}.
\label{Frohlich-12}
\end{eqnarray}
We are in order now to compare our results obtained by means of a non-perturbative method with previous perturbative results. For this we show here the three-phonon self-energy results for the next terms, in the Rayleigh-Schr\"{o}dinger series for $E_0$ and $m^*$~\cite{Dow},

\begin{eqnarray}
E_{0} = -  \omega_0 \left[ \alpha + 0.0159 \alpha^2 + 0.008765 \alpha^3 + {\cal O}(\alpha^4) \right],
\label{Frohlich-13}
\end{eqnarray}

\begin{eqnarray}
m^* = m \left[1 - \frac{\alpha}{6} + 0.02263\alpha^2  + {\cal O}(\alpha^3) \right]^{-1}.
\label{Frohlich-14}
\end{eqnarray}
Notice that there is a relatively good agreement for the effective mass $m^*$, while for the ground state energy, there is a difference in the signal of the coefficient of the $\alpha^2$ term.

\section{Comparison to Related Approaches}
\label{section-DuBois}

In the quantum field theory context, the MSCR resembles the resummation method used by Dolan and Jackiw in Ref.~\cite{T2} to circumvent the problem of breakdown of the perturbative expansion at finite temperature in the $O(N)$ model. The temperature-dependent mass to lowest order is given by

\begin{eqnarray}
m_{\beta}^2 = m^2 + C i \int_k(k^2-m^2)^{-1},
\label{Dolan}
\end{eqnarray}
where $m^2<0$ is a mass parameter introduced to describe symmetry breaking at zero temperature, $C = N \lambda/6$, where $N$ is the number of fields, $\lambda$ is a positive coupling constant, and $\int_k$ is temperature dependent. When higher order terms in the perturbative expansion are added to $m_{\beta}^2$, we have that $m_{\beta}^2$ become negative as the critical temperature is approached from above. To cure this unphysical result, Dolan and Jackiw proceeded with the iteration process

\begin{eqnarray}
m_{\beta}^2 = m^2 + C i \int_k \left[ k^2 -m^2 - C i \int_p(p^2-m^2)^{-1} \right]^{-1},
\label{Dolan2}
\end{eqnarray}
which is equivalent to a summation of daisy diagrams. Continuing the iteration, they were led to a gap-equation for $m_{\beta}^2$

\begin{eqnarray}
m_{\beta}^2 = m^2 + C i \int_k(k^2-m_{\beta}^2)^{-1}.
\label{Dolan3}
\end{eqnarray}
In terms of diagrams, the graphs summed resulting in Eq.~(\ref{Dolan3}) are named as superdaisies~\cite{T2}.

In CM, the MSCR is to some extent very close to the DuBois's approach, i.e., an {\it iteration-perturbation method}, which generates a series of iterations~\cite{Dubois}. To determine the energy $W(p)=\omega(p) + \Sigma(W(p))$ of the physical electron state from the pole of the correct (full) propagator, DuBois employed his iteration-perturbation method, obtaining the solutions as a series of iterations,

\begin{eqnarray}
(0)~~~W&=&\omega,\\
\nonumber
(1)~~~W&=&\omega + {\rm Re} \Sigma(\omega),\\
\nonumber
(2)~~~W&=&\omega + {\rm Re} \Sigma[\omega + \Sigma(\omega)]\\
\nonumber
&=&\omega + {\rm Re} \Sigma(\omega) + {\rm Re} \Sigma'(\omega) {\rm Re} \Sigma(\omega) + ...,
\label{DuBois}
\end{eqnarray}
where $\omega=\omega(p)$, $W=W(p)$, and $\Sigma'(\omega) \equiv \frac{d}{d\omega} \Sigma(\omega) $.

As we mentioned earlier, within the MSCR there is the renormalization of the masses which will run in the loops in all stages (iterations) of the process. This is a basic condition of the approach which guarantees that we will not have divergent masses inside integrals. Otherwise one could have an uncontrollable renormalization process. Another important ingredient of the non-perturbative approach is that it keeps all diagrams at a given order of the perturbative expansion. This assures that the symmetries of the Lagrangian will be respected to a given order of the perturbative expansion~\cite{Caldas1}. 

Several problems faced by different approaches as, for instance, the Hartree, CJT and large-$N$ approximations in the investigation the linear-$\sigma$ model are discussed in Ref.~\cite{Caldas1}.

\section{Summary}
\label{conclusions}

To summarize, we have investigated the effect of non-perturbative corrections to the effective mass in relativistic and in non-relativistic fermion systems. As a worming up, i.e., to present the formalism of our theoretical approach, we have calculated the correction to the electron mass in QED in 3D. Our results of the non-perturbative second-order corrections to the bare electron mass are in good agreement with previous perturbative calculations of order $\alpha^2$ of the perturbative expansion. We have also calculated the electron effective mass in a relativistic Dirac system in 2D, i.e., in graphene, and found the same qualitative behavior found experimentally for the cyclotron mass of charge carriers in graphene, as a function of their concentration $n$.

In the non-relativistic context, we have computed the non-perturbative correction to the effective electron mass due to electron-phonon interaction, and found explicitly the polaron ground state energy and effective mass. Again, we found a relatively good agreement for the effective mass $m^*$ at order $\alpha^2$, whereas for the ground state energy, we found a difference in the signal of the coefficient of the $\alpha^2$ term.

Our approach might also be useful in the investigation of various physical quantities, such as graphene, and other similar fermion systems. One interesting application could also be in the investigation of the percentage of the quark and gluon contributions to the proton mass, which can only be provided by solving QCD phenomenologically~\cite{Ji,Ma} and/or non-perturbatively~\cite{Liu}. 

The study of the non-perturbative corrections to the effective mass and consequently to the pairing gap of imbalanced fermion systems is in progress and will be presented elsewhere.
\vspace{0.5cm}

\section{Acknowledgments}

We would like to thank P. Coleman for stimulating discussions. The author acknowledges partial support by CNPq and FAPEMIG, Brazil.
\vspace{0.1cm}

\appendix

\section{Explicit Calculation of the One-loop Relativistic Electron Self-energy Diagram}
\label{ApA}

In order to evaluate the relativistic electron self-energy, we shall make use of the so called Pauli-Villars regularization. Besides, we modify the photon propagator to avoid difficulties that may arise from infrared divergences in Eq.~(\ref{Sigma}), as

\begin{eqnarray}
\frac{1}{k^2 + i \epsilon} &\to& \frac{1}{k^2 - \lambda^2 + i \epsilon} - \frac{1}{k^2 -\Lambda^2 + i \epsilon}\\
\nonumber
&=& - \int_{\lambda^2}^{\Lambda^2} \frac{dt}{(k^2-t +i \epsilon)^2},
\label{RelativSelf-1}
\end{eqnarray}
where $\lambda$ is a small photon mass (i.e., an infrared cutoff) and $\Lambda$ is an ultraviolet cutoff parameter. QED is restored in limit $\Lambda \to \infty$. Then we have

\begin{widetext}
\begin{eqnarray}
\label{RelativSelf-2}
\Sigma(\slashed{p}) &=&  i e^2  \int \frac{d^4 k}{(2 \pi)^4} \frac{ \gamma^\mu [ \gamma \cdot (p-k) +m ] \gamma_\mu }{(p-k)^2 - m^2}  \int_{\lambda^2}^{\Lambda^2} \frac{dt}{(k^2-t +i \epsilon)^2}\\
\nonumber
&=& \frac{i e^2 }{(2 \pi)^4} \int_{\lambda^2}^{\Lambda^2} dt \int d^4 k \frac{2  (\slashed{k} +m) }{(k^2 - 2 p k + i\epsilon)}  \frac{1}{(k^2-t +i \epsilon)^2}.
\end{eqnarray}
\end{widetext}
We now use the Feynman parametrization on the r.h.s. of Eq.~(\ref{RelativSelf-2})

\begin{eqnarray}
\frac{1}{a^2 b} = 2 \int_0^1 dz \frac{z}{[b+ (a-b)z]^3},
\label{RelativSelf-3}
\end{eqnarray}
to obtain

\begin{eqnarray}
\label{RelativSelf-4}
\Sigma(\slashed{p}) &=& \frac{i e^2 }{(2 \pi)^4} \int_{\lambda^2}^{\Lambda^2} dt \int_0^1 dz \int d^4 k \\
\nonumber
&\times& \frac{4  (\slashed{k} +m) z }{[ k^2 - 2 p k (1-z) -tz +i \epsilon ]^3}.
\end{eqnarray}
The integral in $k$ is solved using the identities

\begin{eqnarray}
\label{RelativSelf-5}
\int \frac{d^4 p}{[ p^2 + 2 p q +t +i \epsilon ]^n} = i \pi^2 \frac{\Gamma(n-2)}{\Gamma(n)} \frac{1}{(t-q^2)^{n-2}},
\end{eqnarray}

\begin{eqnarray}
\label{RelativSelf-6}
\int  \frac{d^4 p~p^{\mu}}{[ p^2 + 2 p q +t +i \epsilon ]^n} = - i \pi^2 \frac{\Gamma(n-2)}{\Gamma(n)} \frac{q^{\mu}}{(t-q^2)^{n-2}},
\end{eqnarray}
where $\Gamma(n)$ is the gamma function. Both results above are valid for $n \geq 3$.

\begin{eqnarray}
\label{RelativSelf-7}
\Sigma(\slashed{p}) &=& \frac{i e^2 }{(2 \pi)^4} \int_0^1 dz  \int_{\lambda^2}^{\Lambda^2} dt   \frac{i 2\pi^2  [(z^2-z)\slashed{p} -m z] }{ tz + p^2(1-z)^2}\\
\nonumber
&=& \frac{ -e^2 }{8 \pi^2} \int_0^1 dz [(z-1)\slashed{p} -m ]z   \int_{\lambda^2}^{\Lambda^2} dt \frac{  1 }{ tz + p^2(1-z)^2}\\
\nonumber
&=& \frac{ e^2 }{8 \pi^2} \int_0^1 dz [(1-z)\slashed{p} + m ]   \frac{ \ln(\Lambda^2 z + p^2(1-z)^2) }{\ln (\lambda^2 z + p^2(1-z)^2)}.\\
\nonumber
\end{eqnarray}
For $\slashed{p} = m$, we have

\begin{eqnarray}
\label{RelativSelf-8}
\Sigma &=&  \frac{ m \alpha }{2 \pi} \int_0^1 dz (2-z)  \frac{ \ln(\Lambda^2 z + m^2(1-z)^2) }{\ln (\lambda^2 z + m^2(1-z)^2)}.\\
\nonumber
\end{eqnarray}
The above expression is not infrared divergent in the limit $\lambda \to 0$, so that we can take $\lambda=0$ in Eq.~(\ref{RelativSelf-8}). Since in the limit $\Lambda \to \infty$, $\Lambda^2 z \gg m^2(1-z)^2$, one obtains~\cite{Itzykson,Milonni}

\begin{eqnarray}
\label{RelativSelf-9}
\Sigma &=&  \frac{ m \alpha }{2 \pi}  \int_0^1 dz (2-z) \left[ \ln \frac{\Lambda^2 }{ m^2}  + \ln \frac{z}{(1-z)^2} \right]\\
\nonumber
&=& \frac{ 3 \alpha m}{4 \pi}  \left[ \ln \frac{\Lambda^2 }{ m^2} +  \frac{1}{2} \right],
\end{eqnarray}
which is the result in Eq.~(\ref{Self1}).

\section{Explicit Calculation of the One-Phonon Electron Self-energy Diagram}
\label{ApC}

The self-energy in Eq.~(\ref{Frohlich-2}) is

\begin{eqnarray}
{\rm Re}\Sigma({\bf k},E) &=& - \frac{M_0^2}{(2 \pi)^3} \int  \frac{d^3 q}{q^2}  \\
\nonumber
&\times& \frac{1}{ \omega_0 -E  + (k^2+q^2 + 2kq \cos \theta)/2m} .
\label{APB-1}
\end{eqnarray}
The three-dimensional integration is $\int d^3 q = \int_0^\infty q^2 dq \int_0^\pi \sin \theta d \theta \int_0^{2 \pi}d\phi = 2 \pi \int_0^\infty q^2 dq \int_{-1}^{1} dy$, where $y=\cos \theta$. Then we have

\begin{eqnarray}
{\rm Re}\Sigma({\bf k},E) &=&  - \frac{M_0^2}{(2 \pi)^2} \int_0^\infty dq \int_{-1}^{1} dy  \\
\nonumber
&\times& \frac{1}{ \omega_0 -E  + (k^2+q^2 + 2kq y)/2m} .
\label{APB-2}
\end{eqnarray}
Making a change of variable from $q$ to $x=\frac{q+k y}{\sqrt{2m}}$, we have

\begin{eqnarray}
{\rm Re}\Sigma({\bf k},E) &=&  - \frac{M_0^2 \sqrt{2m}}{8 \pi^2}  \int_{-1}^{1} dy  \int_{-\infty}^\infty dx\\
\nonumber
&\times& \frac{1}{ \omega_0 -E  + \epsilon_k(1-y^2) + x^2} .
\label{APB-3}
\end{eqnarray}
The integral in $x$ can be done using the formula

\begin{eqnarray}
\int_{-\infty}^{\infty} \frac{d x}{A + x^2} = \frac{\pi \Theta(A)}{\sqrt{A}},
\label{APB-4}
\end{eqnarray}
where $\Theta$ is the step function. In the present case $A = \omega_0 -E  + \epsilon_k(1-y^2) $, and we are left with the $y$ integral,

\begin{eqnarray}
{\rm Re}\Sigma({\bf k},E) &=&  - \frac{M_0^2 \sqrt{2m}}{8 \pi}  \int_{-1}^{1} dy \\
\nonumber
&\times& \frac{1}{ [\omega_0 -E  + \epsilon_k(1-y^2) ]^{1/2} }.
\label{APB-5}
\end{eqnarray}
To perform the integral in $y$, we use the formula

\begin{eqnarray}
\int_{-1}^{1} \frac{d y}{[B- C y^2) ]^{1/2}} = \frac{2}{\sqrt{C}} \sin^{-1} \left( \frac{C}{B} \right)^{1/2},
\label{APB-6}
\end{eqnarray}
where here $B = \omega_0 -E  + \epsilon_k$ and $C = \epsilon_k$, which gives

\begin{eqnarray}
{\rm Re}\Sigma({\bf k},E) &=&  - \frac{ \alpha \omega_0^{3/2}}{\sqrt{\epsilon_k}}  \sin^{-1} \left( \frac{\epsilon_k}{\omega_0 -E  + \epsilon_k} \right)^{1/2}.
\label{APB-7}
\end{eqnarray}
On the mass shell, i.e., making the replacement $E \to \epsilon_k $, the SE is real, and we have

\begin{eqnarray}
\Sigma({\bf k}) &=&  - \frac{ \alpha \omega_0^{3/2}}{\sqrt{\epsilon_k}}  \sin^{-1} \left( \frac{\epsilon_k}{\omega_0} \right)^{1/2}.
\label{APB-8}
\end{eqnarray}
In the limit $\epsilon_k \to 0$, we use the expansion $\sin^{-1}(x)= x + x^3/6 + {\cal O}( x^5)$, and finally obtain~\cite{Mahan}

\begin{eqnarray}
\Sigma({\bf k}) &=&  -  \alpha \omega_0   \left[ 1 + \frac{1}{6} \frac{\epsilon_k}{\omega_0} + {\cal O} \left( \frac{\epsilon_k}{\omega_0} \right)^2  \right],
\label{APB-9}
\end{eqnarray}
which is the result in Eq.~(\ref{Frohlich-3}).

\section{Electron-Phonon Self-energy for $f_{\bf k}$ (Number of) Fermions}
\label{ApD}

Averaging the SE over the Fermi surface allows the introduction of the spectral function $\alpha^2(\omega)F(\omega)$~\cite{Piers},

\begin{eqnarray}
\label{APD-1}
\Sigma(z) &=&  \int_{-\infty}^{\infty} d\epsilon \int_0^{\infty} d\nu \alpha^2(\nu)F(\nu) \\
\nonumber
&\times&\left[ \frac{1+ n(\nu) - f(\epsilon)}{z - ( \epsilon + \nu) } + \frac{ n(\nu) + f(\epsilon)}{z - ( \epsilon - \nu) } \right].
\end{eqnarray}
Taking the zero temperature limit in Eq.~(\ref{APD-1}), the Bose functions $n(\nu)$ vanish and the Fermi functions $f(\epsilon)$ become step functions,

\begin{eqnarray}
\label{APD-2}
\Sigma(z) &=& \int_{-\infty}^{\infty} d\epsilon \int_0^{\infty} d\nu \alpha^2(\nu)F(\nu) \\
\nonumber
&\times& \left[ \frac{\Theta(\epsilon)}{z - ( \epsilon + \nu) } + \frac{ \Theta(-\epsilon)}{z - ( \epsilon - \nu) } \right],
\end{eqnarray}
whose integration in $\epsilon$ yields~\cite{Fetter,Mahan,Piers}

\begin{eqnarray}
\Sigma(z) = \int_0^{\infty}  d \omega \alpha^2(\omega) F(\omega) \ln\left[ \frac{\omega-z}{\omega+z} \right].
\label{Sigma-NonRelat-3}
\end{eqnarray}
In Eq.~(\ref{Sigma-NonRelat-3}) $\alpha^2(\omega) F(\omega)$ is known as the Eliashberg electron-phonon spectral function, where $\alpha(\omega)$ is an effective energy-dependent coupling constant, $F(\omega)$ is the phonon density of states, and $\omega$ is their frequency~\cite{Grimvall}. 

From this expression, one can obtain the dimensionless mass enhancement factor $\lambda$, due to electron-phonon interaction. At low frequencies,

\begin{eqnarray}
\Sigma(\omega) \approx  \Sigma(0) - \lambda \omega,
\label{Sigma-NonRelat-4}
\end{eqnarray}
where

\begin{eqnarray}
\lambda &=& - \frac{d \Sigma (\omega)}{d \omega}|_{\omega=0} \\
\nonumber
&=& 2 \int_0^{\infty}  d \nu \frac{\alpha^2(\nu) F(\nu)}{\nu}.
\label{Sigma-NonRelat-5}
\end{eqnarray}
An analytic expression for $\Sigma(\omega)$ can be derived if, for example, the phonon spectrum is given by the three-dimensional (phenomenological) Debye model. In this case one has~\cite{Piers}

\begin{eqnarray}
\alpha^2(\omega) F(\omega) &=& \lambda \left( \frac{\omega}{\omega_D} \right)^2,~~\rm{if}~~ \omega<\omega_D\\
\nonumber
\alpha^2(\omega) F(\omega) &=& 0,~~\rm{if}~~ \omega>\omega_D,
\label{Sigma-NonRelat-6}
\end{eqnarray}
where $\omega_D$ is the Debye frequency. Plugging $\alpha^2(\omega) F(\omega)$ from the above equation into Eq.~(\ref{Sigma-NonRelat-3}) we obtain

\begin{eqnarray}
\Sigma(z) &=& \int_0^{\omega_D}  d \omega \alpha^2(\omega) F(\omega) \ln\left[ \frac{\omega-z}{\omega+z} \right]\\
\nonumber
&=& \lambda \int_0^{\omega_D}  d \omega  \left( \frac{\omega}{\omega_D} \right)^2 \ln\left[ \frac{\omega-z}{\omega+z} \right]\\
\nonumber
&=& - \lambda \omega_D F\left( \frac{z}{\omega_D} \right),
\label{Sigma-NonRelat-7}
\end{eqnarray}
where

\begin{eqnarray}
F(x) &\equiv&  \frac{1}{3} \left[x+  x^3 \ln \left|1-\frac{1}{x^2} \right| + \ln \left| \frac{1+x}{1-x} \right| \right]\\
\nonumber
&\simeq& x \left[ 1+ \frac{2}{9}x^2 \right],~~~ x\ll1,
\label{Sigma-NonRelat-8}
\end{eqnarray}
where $x= \frac{z}{\omega_D}$. This finally gives~\cite{Sadao}

\begin{eqnarray}
\Sigma(z) &=& - \lambda z  \left[ 1+ \frac{2}{9} \left(\frac{z}{\omega_D} \right)^2 \right],
\label{Sigma-NonRelat-9}
\end{eqnarray}
yielding

\begin{eqnarray}
-\frac{ \partial\Sigma(\vec{k}, \omega)}{\partial \omega}|_{\omega=0} = \lambda = \frac{m^*}{m} - 1,
\label{NREffecM}
\end{eqnarray}
which results in

\begin{eqnarray}
 m^* = (1+ \lambda)m.
\label{NREffecM}
\end{eqnarray}


\begin{thebibliography}{99}



\bibitem{Peskin} See e.g. Chapter 10 of Michael E. Peskin, Daniel V. Schroeder, {\it An Introduction to Quantum Field Theory}, Reading, USA: Addison-Wesley (1995).

\bibitem{Mattuck} Richard D. Mattuck, {\it A Guide to Feynman Diagrams in the Many-Body Problem}, New York, McGraw-Hill (1976).

\bibitem{Grimvall} G. Grimvall, {\it Electron-Phonon Interactions in Metals}, edited by E. P. Wohlfarth, Amsterdam, North-Holland, (1981).

\bibitem{Bohm} H. M. B\"{o}hm, and K. Sch\"{o}rkhuber,  J. Phys.: Condens. Matter {\bf 12}, 2007 (2000).

\bibitem{Krakovsky} A. Krakovsky, and J. K. Percus, Phys. Rev. B {\bf 53}, 7352 (1996).

\bibitem{Vignale} F. G. Eich, Markus Holzmann, and G. Vignale,  Phys. Rev. B {\bf 96}, 035132 (2017).

\bibitem{Scazza} F. Scazza, G. Valtolina, P. Massignan, A. Recati, A. Amico, A. Burchianti, C. Fort, M. Inguscio, M. Zaccanti, and G. Roati, Phys. Rev. Lett. {\bf 118}, 083602 (2017).

\bibitem{Cui} W. Li, and X. Cui, Phys. Rev. A {\bf 96}, 053609 (2017).

\bibitem{T1} D. A. Kirzhnits and A. D. Linde, Phys. Lett. {\bf B42}, 471 (1972); Ann. Phys. {\bf 101}, 195 (1976).

\bibitem{T2} L. Dolan and R. Jackiw, Phys. Rev. D {\bf 9}, 3320 (1974).

\bibitem{T3} S. Weinberg, Phys. Rev. D {\bf 9}, 3357 (1974).

\bibitem{MMarino} M. Mari\~{n}o, {\it Instantons and Large N: An Introduction to Non-Perturbative Methods in Quantum Field Theory}, Cambridge, Cambridge University Press (2015).

\bibitem{Haken} H. Haken, H. C. Wolf, {\it The Physics of Atoms and Quanta: Introduction to Experiments and Theory}, Berlin Heidelberg, Springer-Verlag (2005).

\bibitem{Lamb1} Th. St\"{o}hlker, P. H. Mokler, K. Beckert, F. Bosch, H. Eickhoff, B. Franzke, M. Jung, T. Kandler, O. Klepper, C. Kozhuharov, R. Moshammer, F. Nolden, H. Reich, P. Rymuza, P. Sp\"{a}dtke, and M. Steck, Phys. Rev. Lett. {\bf 71}, 2184 (1993).

\bibitem{Lamb2} H. F. Beyer, D. Liesen, F. Bosch, K. D. Finlayson, M. Jung, O. Klepper, R. Moshammer, K. Beckert, B. Franke, F. Nolden, P. Sp\"{a}dtke, M. Steck, G. Menzel, and D. R. Deslattes, Phys. Lett. A {\bf 184}, 435 (1994).

\bibitem{Michele} Michele Maggiore, { \it A Modern Introduction to Quantum Field Theory}, New York, Oxford University Press (2005).

\bibitem{Weisskopf} V. Weisskopf, Phys. Rev. {\bf 56}, 72 (1939).

\bibitem{Caldas1} H. Caldas, A. L. Mota and M. C. Nemes, Phys. Rev. D {\bf 63}, 56011 (2001).

\bibitem{Caldas2} H. Caldas, Phys. Rev. D {\bf 65}, 65005 (2002).

\bibitem{Caldas3} H. Caldas, Phys. Rev. D {\bf 66}, 105015 (2002).

\bibitem{Leonti} L. Labzowsky, A. Mitrushenkov, V. Shelyuto, and G. Soff, Phys. Rev. A {\bf 97}, 4038 (1998).

\bibitem{Leonti2} L. Labzowsky, A. Mitrushenkov, V. Shelyuto, and G. Soff, Phys. Lett. A {\bf 240}, 225 (1998).

\bibitem{Hanneke} D. Hanneke, S. Fogwell, and G. Gabrielse,  Phys. Rev. Lett. {\bf 100}, 120801 (2008).

\bibitem{Sturm} S. Sturm et al. Nature {\bf 506}, 467 (2014).

\bibitem{Bakulev} A. P. Bakulev, and D. V. Shirkov, arXiv:1102.2380 (2011).

\bibitem{Kotov} V. N. Kotov, B. Uchoa, V. M. Pereira, F. Guinea, and A. H. Castro Neto, Rev. Mod. Phys. {\bf 84}, 1067 (2012).

\bibitem{Das} E. Barnes, E. H. Hwang, R. E. Throckmorton, and S. Das Sarma, Phys. Rev. B {\bf 89}, 235431 (2014).

\bibitem{Novoselov} K. S. Novoselov, A. K. Geim, S. V. Morozov, D. Jiang, M. I. Katsnelson, I. V. Grigorieva, S. V. Dubonos, and A. A. Firsov, Nature {\bf 438}, 197 (2005).

\bibitem{Zhang} Y. Zhang, Y.-W. Tan, H. L. Stormer, and P. Kim, Nature {\bf 438}, 201 (2005).

\bibitem{PhysB} H. Caldas, Physica B: Condensed Matter {\bf 577}, 411814 (2020).

\bibitem{Elias} D. C. Elias, R. V. Gorbachev, A. S. Mayorov, S. V. Morozov, A. A. Zhukov, P. Blake, L. A. Ponomarenko, I. V. Grigorieva, K. S. Novoselov, F. Guinea, A. K. Geim, Nat. Phys. {\bf 7} 701 (2011).

\bibitem{Piers} P. Coleman, {\it Introduction to Many-body Physics}, Cambridge, Cambridge University Press (2016).

\bibitem{Frohlich} H. Fr\"{o}hlich, {\it Introduction to the theory of the polaron}. In: {\it Polarons and Excitons}, ed. C. G. Kuper and G. D. Whitfield, New York, Plenum Press (1963), pp. 1-32.

\bibitem{Daniel} D. I. Khomskii, {\it Basic Aspects of the Quantum Theory of Solids: Order and Elementary Excitations}, Cambridge, Cambridge University Press (2010).

\bibitem{Fetter} A. L. Fetter and J. D. Walecka, {\it Quantum Theory of Many-Particle Systems}, San Francisco, McGraw-Hill (1971).

\bibitem{Mahan} G. D. Mahan, {\it Many-Particle Physics}, New York, Plenum Press (1990).

\bibitem{Giuliani} G. F. Giuliani and G. Vignale, {\it Quantum Theory of the Electron Liquid}, chapter 8, Cambridge University Press (2005).

\bibitem{Dow} P. Sheng and J. D. Dow, Phys. Rev. B {\bf 4}, 1343 (1971).

\bibitem{Dubois} D. F. DuBois, Ann. Phys. (N. Y.) {\bf 7}, 174 (1959); 8, 24 (1959).

\bibitem{Itzykson} C. Itzykson and J.-B. Zuber, {\it Quantum field theory}, New York, McGraw-Hill (1985).

\bibitem{Milonni} P. W. Milonni, {\it The Quantum Vacuum. An introduction to Quantum Electrodynamics}, San Diego, Academic Press (1993), pp. 394-405.

\bibitem{Sadao} S. Nakajima and M. Watabe, Progr. Theoret. Phys. (Kyoto) {\bf 29}, 341 (1963).

\bibitem{Ji} X.-D. Ji, Phys. Rev. Lett. {\bf 74}, 1071 (1995).

\bibitem{Ma} Y.-B. Yang, Y. Chen, T. Draper, M. Gong, K.-F. Liu, Z. Liu, and J.-P. Ma, Phys. Rev. D {\bf 91}, 074516 (2015).

\bibitem{Liu} Y.-B. Yang, J. Lian g, Y. J. Bi, Y. Chen, T. Draper, K.-F. Liu and Z. Liu, Phys. Rev. Lett. {\bf 121}, 212001 (2018).


\end{thebibliography}
\end{document}